\documentclass{article}
\usepackage{amssymb}

\usepackage{graphicx}
\usepackage{amsmath}
\usepackage{epsfig}

\newtheorem{theorem}{Theorem}

\newtheorem{proposition}[theorem]{Proposition}

\begin{document}

\title{Bubble effect on Kelvin-Helmholtz' instability }
\author{S. L. Gavrilyuk$^{\ast }$, H. Gouin\thanks{Laboratoire de Mod\'{e}lisation en M\'{e}canique et Thermodynamique,
Facult\'{e} des Sciences et Techniques de Saint-J\'{e}r\^{o}me, Case 322,
Avenue Escadrille Normandie-Niemen, 13397 Marseille Cedex 20, France,
e-mail: sergey.gavrilyuk@univ.u-3mrs.fr; henri.gouin@univ.u-3mrs.fr} \ and
V.M.Teshukov\thanks{
Lavrentyev Institute of Hydrodynamics, Lavrentyev prospect 15, Novosibirsk
630090, Russia, also INRIA,\ SMASH, 2004 route de Lucioles, 06902 Sophia
Antipolis, France, e-mail: teshukov@hydro.nsc.ru}}
\maketitle

\begin{abstract}
We derive boundary conditions at interfaces (contact
discontinuities) for a class of Lagrangian models describing, in
particular, bubbly flows. We use these conditions to study
Kelvin-Helmholtz' instability which develops in the flow of two
superposed layers of a pure incompressible fluid and a fluid
containing gas bubbles, co-flowing with different velocities. We
show that the presence of bubbles in one layer stabilizes the flow
in some intervals of wave lengths.
\end{abstract}

\section{Introduction}

Many mathematical models of fluid mechanics are derived through
the approximation of the solution of boundary value problems for
Euler equations. For example, equations for bubbly flows are
derived as an approximation of the solution of a complex
free-boundary problem describing the motion of the mixture of
water and gas bubbles (Iordanski (1960), Kogarko (1961),
Wijngaarden (1968)). Green-Naghdi's model describing wave motions
of a liquid layer of finite depth in a shallow water approximation
with account of dispersion effects is another example (Green
\textit{et al} (1974), Green \& Naghdi (1976)). Due to the fact
that averaging procedures and asymptotic expansions have been used
in the derivation, it is not obvious to decide what boundary
conditions are natural for these systems of equations. In this
paper, a Hamiltonian  formulation of the problem in
($t,\mathbf{x)}$-space is proposed which allows one to find
boundary conditions for a general class of models. This class
includes a model of bubbly fluid and dispersive shallow water. We
use the boundary conditions to study the stability of two
co-flowing layers of bubbly and pure fluids. In absence of gravity
and capillarity, it is well known that Kelvin-Helmholtz'
instability develops for any wave lengths in the flow of two
superposed layers of pure incompressible fluids  (see, for
example, Drazin \& Reid (1981)). We prove that the presence of
bubbles in one layer can produce a stabilizing effect: the flow
becomes stable with respect to perturbations of some wave lengths.

\section{Governing equations and conditions on moving interfaces}

\setcounter{equation}{0}
\subsection{Variation of Hamilton's action}
Here we calculate the variation of Hamilton's action for a special
class of Lagrangians. Usually this procedure is used only for
derivation of governing equations.  We  obtain below not only
governing equations but also boundary conditions at inner contact
surfaces.

Let us consider the Lagrangian of the form:
\begin{equation}
L=L(\mathbf{J},\frac{\partial \mathbf{J}}{\partial \mathbf{z}},\mathbf{z})
\label{lagr}
\end{equation}
where $\mathbf{z=}\left(
\begin{array}{c}
t \\
\mathbf{x}
\end{array}
\right) \equiv \left( z^{i}\right) ,$ $(i=0,1,2,3),\ t$ is the time, $%
\mathbf{x}$ is the space variable, $\mathbf{J=}\left(
\begin{array}{c}
\rho \\
\rho \mathbf{u}
\end{array}
\right) ,$ $\rho $ is the fluid density and $\mathbf{u}$ is the
velocity field\textbf{.} Let us calculate the variation of
Hamilton's action in the case when the 4-D momentum $\mathbf{J}$
verifies the equation of continuity:
\begin{equation}
Div\text{ }\mathbf{J}\equiv \dfrac{\partial \rho }{\partial t}+div(\rho
\mathbf{u})=0  \label{constr}
\end{equation}
Hamilton's action is defined by
\begin{equation}
a=\int_{\Omega }Ld\mathbf{z}  \label{action}
\end{equation}
where $\Omega $ is a material domain. Considering a smooth
one-parameter family of virtual motions
\begin{equation*}
{\mathbf{z}}=\mathbf{\Phi }(\mathbf{Z},\varepsilon ),\quad \mathbf{\Phi }(%
\mathbf{Z},0)={\mbox{{\boldmath $\varphi$}}}(\mathbf{Z})
\end{equation*}
($\mathbf{Z}$\ stands for the Lagrangian coordinates,
$\varepsilon$ is a small parameter at the vicinity of zero and
$\mathbf{z}={\mbox{\boldmath $\varphi$}}(\mathbf{Z})$ is the real
motion),  we define the virtual displacements ${\mbox{\boldmath
$\zeta$}}(\mathbf{Z})$ and the Lagrangian variations $\delta
\mathbf{J(Z)}$  by the formulae:
\begin{equation}
{\mbox{\boldmath $\zeta$}}({\mathbf Z})=\left. \frac{\partial
\mathbf{\Phi}({\mathbf Z},\varepsilon)}{\partial \varepsilon
}\right|_{\varepsilon =0},\quad \delta
\mathbf{J(Z)=}\left. \frac{\partial \mathbf{J(Z,}\varepsilon \mathbf{)}}{%
\partial \varepsilon }\right| _{\varepsilon =0}  \label{variation}
\end{equation}
Due to the fact that $\mathbf{Z}={\mbox{\boldmath
$\varphi$}}^{-1}\mathbf{(z)}$, we consider the variations as
functions of Eulerian coordinates and use the same notations
${\mbox{\boldmath $\zeta$}}(\mathbf{z})$ and $\delta
\mathbf{J(z)}$ in Eulerian variables. Hamilton's principle assumes
that ${\mbox{\boldmath $\zeta$}}(\mathbf{z})=0$ on the boundary
$\partial \Omega $ of $\Omega $.
\newline
In the following,  the transposition is denoted by $^{T}$. For any
vectors $\mathbf{a}$ and $\mathbf{b}$ we use the notation
$\mathbf{a}^{T}\mathbf{b}$ for their scalar product
$\mathbf{a}\cdot \mathbf{b}$ and $\mathbf{ab}^{T}$ for their
tensor product $\mathbf{a}\otimes \mathbf{b}$. The divergence of
the second order tensor $A$ is a covector defined by
\begin{equation*}
Div\left( A\mathbf{h}\right) =Div\left( A\right) \mathbf{h}
\end{equation*}
where $\mathbf{h}$ is any constant vector field. $Div$ and $div$,
$Grad$ and $\nabla$ are respectively divergence and gradient
operators in the \textit{4-D} and \textit{3-D} space. The identity
matrix and the zero matrix of dimension $n$ are denoted by $I_{n}$
and $O_{n}$.\newline In calculations we shall use the equality
\begin{equation}
\delta \mathbf{J=}\left( \frac{\partial {\mbox{\boldmath
$\zeta$}}}{\partial \mathbf{z}} -\left( Div\text{
}{\mbox{\boldmath $\zeta$}}\right) \text{ }I_{4}\right) \mathbf{J}
\label{moment}
\end{equation}
which was proved in Gavrilyuk \& Gouin (1999). Variation of the
Hamilton action is
\begin{equation}
\delta a=\left. \frac{da}{d\varepsilon }\right| _{\varepsilon
=0}=\int_{\Omega }\left( \delta L+L\text{ }tr\left( \frac{\partial
{\mbox{\boldmath $\zeta$}}}{\partial \mathbf{z}}\right) \right)
d\mathbf{z}  \label{var1}
\end{equation}
Since
\begin{equation*}
\delta L=\frac{\partial L}{\partial \mathbf{J}}\ \delta
\mathbf{J+}tr\left( \frac{\partial L}{\partial \left(
\dfrac{\partial \mathbf{J}}{\partial \mathbf{z}}\right) }\delta
\left( \frac{\partial \mathbf{J}}{\partial \mathbf{z}}\right)
\right) +\frac{\partial L}{\partial \mathbf{z}}\ {\mbox{\boldmath
$\zeta$}}
\end{equation*}
and
\begin{equation*}
\delta \left( \frac{\partial \mathbf{J}}{\partial \mathbf{z}}\right) =\frac{%
\partial \delta \mathbf{J}}{\partial \mathbf{z}}-\frac{\partial \mathbf{J}}{%
\partial \mathbf{z}}\frac{\partial {\mbox{\boldmath $\zeta$}}}{\partial \mathbf{z}}
\end{equation*}
we get from (\ref{moment}) and (\ref{var1}):
\begin{equation*}
\delta a=\int_{\Omega }\left( \frac{\partial L}{\partial
\mathbf{J}}\left( \frac{\partial {\mbox{\boldmath
$\zeta$}}}{\partial \mathbf{z}}-\left( Div\ {\mbox{\boldmath
$\zeta$}}\right) I_{4}\right) \mathbf{J+}tr\left( A^{T}\left(
\frac{\partial \delta \mathbf{J}}{\partial
\mathbf{z}}-\frac{\partial \mathbf{J}}{\partial
\mathbf{z}}\frac{\partial {\mbox{\boldmath $\zeta$}}}{\partial
\mathbf{z}}\right) \right) +\frac{\partial L}{\partial
\mathbf{z}}\ {\mbox{\boldmath $\zeta$}}+L\ Div\ {\mbox{\boldmath
$\zeta$}}\right) d\mathbf{z}
\end{equation*}
with
\begin{equation}
A^{T}=\frac{\partial L}{\partial \left( \dfrac{\partial \mathbf{J}}{\partial
\mathbf{z}}\right) }\text{ or }\left( A^{T}\right) _{i}^{j}=\frac{\partial L%
}{\left( \dfrac{\partial J^{i}}{\partial z^{j}}\right) }  \label{matrix}
\end{equation}
For the sake of simplicity the measure of integration will not be
indicated. Since for any linear transformation $A$ and vector
field $\mathbf{v}$
\begin{equation*}
Div(A\mathbf{v})=(Div\text{ }A)\mathbf{v+}tr(A\dfrac{\partial
\mathbf{v}}{\partial \mathbf{z}})
\end{equation*}
we get
\begin{equation*}
\delta a=\int_{\Omega }tr\left( \left( \mathbf{J}\frac{\partial
L}{\partial \mathbf{J}}-A^{T}\frac{\partial \mathbf{J}}{\partial
\mathbf{z}}\right) \frac{\partial {\mbox{\boldmath
$\zeta$}}}{\partial \mathbf{z}}\right) +\left( L-\frac{\partial
L}{\partial \mathbf{J}}\ \mathbf{J}\right) Div\ {\mbox{\boldmath
$\zeta$}}+tr\left( A^{T}\frac{\partial \delta \mathbf{J}}{\partial
\mathbf{z}}\right) +\frac{\partial L}{\partial \mathbf{z}}\
{\mbox{\boldmath $\zeta$}}=
\end{equation*}
\begin{equation*}
=\int_{\Omega }tr\left( \left( \mathbf{J}\frac{\partial
L}{\partial \mathbf{J}}-A^{T}\frac{\partial \mathbf{J}}{\partial
\mathbf{z}}\right) \frac{\partial {\mbox{\boldmath
$\zeta$}}}{\partial \mathbf{z}}\right) \mathbf{+}\left( L-
\frac{\partial L}{\partial \mathbf{J}}\ \mathbf{J}\right) Div\
{\mbox{\boldmath $\zeta$}}+
\end{equation*}
\begin{equation*}
\mathbf{+}Div\left( A^{T}\delta \mathbf{J}\right) -Div\left(
A^{T}\right) \delta \mathbf{J}+\frac{\partial L}{\partial
\mathbf{z}}\ {\mbox{\boldmath $\zeta$}}\ ,
\end{equation*}
and finally
\begin{equation*}
\delta a=\int_{\Omega }tr\left( \left( \mathbf{J}\frac{\partial
L}{\partial \mathbf{J}}-A^{T}\frac{\partial \mathbf{J}}{\partial
\mathbf{z}}\right) \frac{\partial {\mbox{\boldmath
$\zeta$}}}{\partial \mathbf{z}}\right) \mathbf{+}\left(
L-\frac{\partial L}{\partial \mathbf{J}}\ \mathbf{J}\right) Div\
{\mbox{\boldmath $\zeta$}}+Div\left( A^{T}\delta \mathbf{J}\right)
-
\end{equation*}
\begin{equation*}
-Div\left( A^{T}\right) \left( \frac{\partial {\mbox{\boldmath
$\zeta$}}}{\partial \mathbf{z}}-\left( Div\ {\mbox{\boldmath
$\zeta$}}\right) I_{4}\right) \mathbf{J}+\frac{\partial
L}{\partial \mathbf{z}}\ {\mbox{\boldmath $\zeta$}}=
\end{equation*}
\begin{equation*}
=\int_{\Omega }tr\left( \left( \mathbf{J}\frac{\partial
L}{\partial \mathbf{J}}-A^{T}\frac{\partial \mathbf{J}}{\partial
\mathbf{z}}-\mathbf{J}Div\left( A^{T}\right) \right)
\frac{\partial {\mbox{\boldmath $\zeta$}}}{\partial \mathbf{z}}
\right) +
\end{equation*}
\begin{equation*}
+\left( L-\frac{\partial L}{\partial \mathbf{J}}\
\mathbf{J}+\left( DivA^{T}\right) \mathbf{J}\right) Div\
{\mbox{\boldmath $\zeta$}}+Div\left( A^{T}\delta \mathbf{J}\right)
+\frac{\partial L}{\partial \mathbf{z}}\ {\mbox{\boldmath
$\zeta$}}=
\end{equation*}
\begin{equation*}
=\int_{\Omega }Div\left( \left( \mathbf{J}\frac{\partial
L}{\partial \mathbf{J}}-A^{T}\frac{\partial \mathbf{J}}{\partial
\mathbf{z}}-\mathbf{J}\
Div\left( A^{T}\right) \right) \ {\mbox{\boldmath $\zeta$}} +A^{T}\delta \mathbf{J}%
\right) -
\end{equation*}
\begin{equation*}
-Div\left( \mathbf{J}\frac{\partial L}{\partial \mathbf{J}}-A^{T}\frac{%
\partial \mathbf{J}}{\partial \mathbf{z}}-\mathbf{J}\text{ }Div\left(
A^{T}\right) \right) \ {\mbox{\boldmath $\zeta$}}+
\end{equation*}
\begin{equation*}
+Div\left( \left( L-\frac{\partial L}{\partial
\mathbf{J}}\mathbf{J+}\left( DivA^{T}\right) \mathbf{J}\right) \
{\mbox{\boldmath $\zeta$}}\right) -Grad\left( L- \frac{\partial
L}{\partial \mathbf{J}}\mathbf{J+}\left( DivA^{T}\right)
\mathbf{J}\right) ^{T}\ {\mbox{\boldmath $\zeta$}}+\frac{\partial
L}{\partial \mathbf{z}}\ {\mbox{\boldmath $\zeta$}}
\end{equation*}
Let us denote
\begin{equation*}
\frac{\delta L}{\delta \mathbf{J}}\equiv \frac{\partial
L}{\partial \mathbf{J}}- Div\left( A^{T}\right) =\mathbf{K}^{T}.
\end{equation*}
Then
\begin{equation}
\delta a=\int_{\Omega }\left( \frac{\partial L}{\partial \mathbf{z}}%
-Div\left( \mathbf{J}\text{ }\mathbf{K}^{T}-A^{T}\frac{\partial \mathbf{J}}{%
\partial \mathbf{z}}+\left( L-\mathbf{K}^{T}\ \mathbf{J}\right) \text{ }%
I_{4}\right) \right) \ {\mbox{\boldmath $\zeta$}}+
\label{actionvar}
\end{equation}
\begin{equation*}
+\int_{\partial \Omega }{\mathbf{N}^{T}}\left( \mathbf{J}\
\mathbf{K}
^{T}-A^{T}\frac{\partial \mathbf{J}}{\partial \mathbf{z}}+\left( L-\mathbf{K}%
^{T}\ \mathbf{J}\right) \text{ }I_{4}\right){\mbox{\boldmath
$\zeta$}}+{\mathbf{N}^{T}}A^{T}\delta \mathbf{J}
\end{equation*}
with $\mathbf{N}^{T}=(-D_{n},\mathbf{n}^{T})$, $D_{n}$ denotes the
surface velocity and $\mathbf{n}$ is the space unit normal vector.
Virtual displacements vanish at the boundary $\partial \Omega $
and the surface integral is zero. The volume integral yields the
equations of motion in conservative form as in Gavrilyuk \& Gouin
(1999).
\newline
In the case when fluid tensorial quantities are discontinuous at
the inner interface $\Sigma $, expression (\ref{actionvar})
becomes
\begin{equation}
\delta a=\int_{\Omega }\left( \frac{\partial L}{\partial \mathbf{z}}%
-Div\left( \mathbf{J}\text{ }\mathbf{K}^{T}-A^{T}\frac{\partial \mathbf{J}}{%
\partial \mathbf{z}}+\left( L-\mathbf{K}^{T}\ \mathbf{J}\right) \text{ }%
I_{4}\right) \right) \ {\mbox{\boldmath $\zeta$}}+
\label{actionvar1}
\end{equation}
\begin{equation*}
+\int_{\Sigma }\left[ {\mathbf{N}^{T}}\left( \mathbf{J}\text{ }\mathbf{K}%
^{T}-A^{T}\frac{\partial \mathbf{J}}{\partial \mathbf{z}}+\left( L-\mathbf{K}%
^{T}\ \mathbf{J}\right) \text{ }I_{4}\right) \ {\mbox{\boldmath
$\zeta$}}+{\mathbf{N}^{T}}A^{T}\delta \mathbf{J}\right]
\end{equation*}
where the jump through $\Sigma $ is denoted by $\left[ \quad
\right] $. \newline
\subsection{Governing equations and boundary conditions}
We explicit the governing equations and the inner boundary
conditions for the Lagrangian
\begin{equation}
L=\frac{1}{2}\rho \left| \mathbf{u}\right| ^{2}-W\left( \rho ,\overset{\cdot
}{\rho }\right) ,\quad \text{where \ }\overset{\cdot }{\left( \;\right) \;}=%
\dfrac{d}{dt}=\dfrac{\partial \left( \;\right) }{\partial
t}+{\mathbf{u}}^T {\mbox{\boldmath $\nabla$}} \left( \;\right)
\label{ll}
\end{equation}
Such a Lagrangian appears in the study of wave propagation in both
shallow water flows with dispersion and bubbly flows (a complete
discussion of these models is given in Gavrilyuk \& Teshukov
(2001)). We get
\begin{equation*}
\frac{\partial \mathbf{J}}{\partial \mathbf{z}}=\left(
\begin{array}{ll}
\;\dfrac{\partial \rho }{\partial t} & \;\dfrac{\partial \rho }{\partial
\mathbf{x}} \\
\begin{array}{l}
\\
\dfrac{\partial \mathbf{j}}{\partial t}
\end{array}
&
\begin{array}{l}
\\
\dfrac{\partial \mathbf{j}}{\partial \mathbf{x}}
\end{array}
\end{array}
\right) ,\quad \frac{\partial L}{\partial \mathbf{J}}=\left( -\frac{\left|
\mathbf{u}\right| ^{2}}{2}-\dfrac{\partial W}{\partial \rho }+\frac{1}{\rho }%
\dfrac{\partial W}{\partial \overset{\cdot }{\rho }}\left( \mathbf{\nabla }%
\rho \right) ^{T}\mathbf{u,}\text{ }\mathbf{u}^{T}-\frac{1}{\rho }\dfrac{%
\partial W}{\partial \overset{\cdot }{\rho }}\left( \mathbf{\nabla }\rho
\right) ^{T}\right)
\end{equation*}
Matrix (\ref{matrix}) becomes
\begin{equation*}
A^{T}=\frac{\partial L}{\partial \left( \dfrac{\partial \mathbf{J}}{\partial
\mathbf{z}}\right) }=\left(
\begin{array}{ll}
\;-\dfrac{\partial W}{\partial \overset{\cdot }{\rho }} & \;\mathbf{0}^{T}
\\
\begin{array}{l}
\\
-\dfrac{\partial W}{\partial \overset{\cdot }{\rho }}\mathbf{u}
\end{array}
&
\begin{array}{l}
\\
O_{3}
\end{array}
\end{array}
\right) .
\end{equation*}
\newline
We get
\begin{equation}
Div(A^{T})=\left( -\dfrac{\partial }{\partial t}\left( \dfrac{\partial W}{%
\partial \overset{\cdot }{\rho }}\right) -div\left( \dfrac{\partial W}{%
\partial \overset{\cdot }{\rho }}\mathbf{u}\right) ,\;\mathbf{0}^{T}\right) ,
\label{formulae}
\end{equation}
\begin{equation*}
\mathbf{K}^{T}\mathbf{=}\left( -\frac{\left| \mathbf{u}\right|
^{2}}{2}-\dfrac{\delta W}{\delta \rho
}+\frac{1}{\rho}\dfrac{\partial W}{\partial
\overset{\cdot }{\rho }}\left( \mathbf{\nabla }\rho \right) ^{T}\mathbf{u,}%
\text{ }\mathbf{u}^{T}-\frac{1}{\rho }\dfrac{\partial W}{\partial \overset{%
\cdot }{\rho }}\left( \mathbf{\nabla }\rho \right) ^{T}\right),
\end{equation*}
\begin{equation*}
\mathbf{N}^{T}\ \mathbf{J=\;}\rho \left( \mathbf{n}^{T}\
\mathbf{u}- D_{n}\right) ,\quad \mathbf{N}^{T}A^{T}=\left(
-\dfrac{\partial W}{\partial \overset{\cdot }{\rho }}\left(
\mathbf{n}^{T}\ \mathbf{u-}D_{n}\right), \mathbf{0}^{T}\right).
\end{equation*}
We study the case when $\Sigma$ is a contact  surface and,
consequently, $\mathbf{N}^{T}\ \mathbf{J}=0$,
$\mathbf{N}^{T}A^{T}=\mathbf{0}^{T}$. We see that the surface
integral in  (\ref{actionvar1}) vanishes if the boundary
conditions :
\begin{equation}
\left[ L-\mathbf{K}^{T}\ \mathbf{J}\right] \equiv\left[ p\right]
=0 \label{bondcond}
\end{equation}
are fulfilled with $p\equiv L-\mathbf{K}^{T}\ \mathbf{J}$.
Vanishing the volume integral in (\ref{actionvar1}) and using
relations (\ref{formulae}), we obtain the governing equations in
the form
\begin{equation}
\dfrac{\partial \rho }{\partial t}+div(\rho \mathbf{u})=0,
\label{sysnon}
\end{equation}
\begin{equation*}
\dfrac{\partial \rho \mathbf{u}}{\partial t}+div\left( \rho \mathbf{uu}^{T}%
\mathbf{+}pI\right) =0,
\end{equation*}
\begin{equation*}
p\equiv \rho \left( \dfrac{\partial W}{\partial \rho }-\dfrac{\partial }{%
\partial t}\left( \dfrac{\partial W}{\partial \overset{\cdot }{\rho }}%
\right) -div\left( \dfrac{\partial W}{\partial \overset{\cdot }{\rho }}%
\mathbf{u}\right) \right) -W =\rho\frac{\delta W}{\delta \rho}-W.
\end{equation*}
A complete set of boundary  conditions  at the contact interfaces
for the class of models  considered is:
\begin{equation}
\mathbf{n}^{T}\mathbf{u-}D_{n}=0,\quad \left[ p\right] =0
\label{pressure}
\end{equation}
Equations (\ref{sysnon}) have been obtained earlier in Gavrilyuk
\& Shugrin (1996) by means of the classical method of Lagrange
multipliers. Notice that this method did not give jump conditions
(\ref{bondcond}). System (\ref {sysnon}) is reminiscent of Euler
equations for compressible fluids. The term $p$ in equations of
motion (\ref{sysnon}) and boundary conditions (\ref{pressure})
stands for the pressure. Nevertheless, $p$ is not a function  of
density as in the case of barotropic fluids: it depends also on
material time derivatives of the density. Thus, $p$ is not usual
thermodynamic pressure. In general, it implies that the boundary
conditions are not just consequences of conservation laws.   For
example,  for Korteweg-type fluids where the stress tensor depends
on the density gradient, the boundary conditions contain normal
derivatives of the density at interfaces (Seppecher (1989), Gouin
\& Gavrilyuk (1999)). As we have shown here, the boundary
conditions at inner interfaces  in bubbly fluids do not depend on
normal derivatives of the density: they involve only tangential
derivatives.\newline If the flow domain is shared between two
domains consisting of pure ideal fluid and bubbly fluid, all
previous calculations are still valid and boundary conditions
(\ref{pressure}) can be extended to such a case.

\subsection{Exact statement of the problem}
Now, we propose to use conditions (\ref{pressure}) to study
Kelvin-Helmholtz' instability of a parallel flow between two rigid
walls of a layer of  bubbly fluid in contact with a layer of pure
incompressible fluid.

Consider the 2-D flow of two superposed finite layers of inviscid
fluids in the channel: $-\infty < x < +\infty$, $0 < y < H_{0}$.
The first one is governed by the Euler equations :
\begin{equation}
div(\mathbf{u}_{1})=0,\quad \frac{\partial \mathbf{u}_{1}}{\partial t}%
+\left( \mathbf{u}_{1}^{T}\mathbf{\nabla }\right) \mathbf{u}_{1}+\dfrac{%
\mathbf{\nabla }p_{1}}{\rho _{10}}=0,\quad -\infty <x<\infty ,\text{ }%
0<y<h(t,x)  \label{pure}
\end{equation}
Here $\rho _{10}$ is the constant density of pure fluid,
$\mathbf{u}_{1}=(u_{1},v_{1})^T$ is the velocity field and $p_{1}$ is the pressure.%
\newline
In the domain $h(t,x)<y<H_{0}$ governing equations (\ref{sysnon})
for the second fluid are
\begin{equation}
\dfrac{\partial \rho _{2}}{\partial t}+div(\rho
_{2}\mathbf{u}_{2})=0, \label{bubbly}
\end{equation}
\begin{equation*}
\frac{\partial \mathbf{u}_{2}}{\partial t}+\left( \mathbf{u}_{2}^{T}\mathbf{%
\nabla }\right) \mathbf{u}_{2}+\dfrac{\mathbf{\nabla }p_{2}}{\rho
_{2}}=0,
\end{equation*}
\begin{equation*}
p_{2}=\rho _{2}\left( \dfrac{\partial W}{\partial \rho
_{2}}-\dfrac{\partial
}{\partial t}\left( \dfrac{\partial W}{\partial \overset{\cdot }{\rho }_{2}}%
\right) -div\left( \dfrac{\partial W}{\partial \overset{\cdot }{\rho }_{2}}%
\mathbf{u}_{2}\right) \right) -W,\quad W=W(\rho _{2},\overset{\cdot }{\rho }%
_{2})
\end{equation*}
Here  $\rho _{2}$, $\mathbf{u}_{2}=(u_{2},v_{2})^T$ and $p_{2}$
are the average density, velocity and mixture pressure,
respectively; $W=W(\rho _{2},\overset{\cdot }{\rho }_{2})$  is a
given potential. For the flow of compressible bubbles having the
same radius, potential $W$ has the form (see, for example,
Gavrilyuk (1994) and Gavrilyuk \& Teshukov (2001)):
\begin{equation}
W(\rho _{2},\overset{\cdot }{\rho
}_{2})=\rho_2\left(c_{g}\varepsilon _{g}(\rho _{g})-2\pi n\rho
_{l}R^{3}\overset{\cdot }{R}^{2}\right) \label{potential}
\end{equation}
where $\varepsilon _{g}$ is the internal energy of the gas in
bubbles, $R$ is the bubble radius, $\rho _{g}$ is the gas density,
$\rho_{l}=const$ is the density of carrying phase, $c_{g}=const$
is the bubble mass concentration, $n$ is the number of bubbles per
unit mass. Potential $W$ is the difference between the internal
energy of gas and the kinetic energy of fluid due to radial bubble
oscillations. The bubble radius
and the bubble density are functions of the average density $\rho _{2}:$%
\begin{equation*}
\dfrac{4}{3}\pi R^{3}=\frac{1}{n}\left( \dfrac{1}{\rho _{2}}-\dfrac{1-c_{g}}{%
\rho _{l}}\right) ,\quad \rho _{g}=c_{g}\left( \dfrac{1}{\rho _{2}}-\dfrac{%
1-c_{g}}{\rho _{l}}\right) ^{-1}
\end{equation*}
It can be shown that governing equations (\ref{bubbly}) with
potential (\ref{potential}) coincide exactly with classical
equations of bubbly fluids (Kogarko (1961) and van Wijngaarden
(1968)). In particular, the role of "equation of state"
\begin{equation}
p_{2}=\rho _{2} \frac{\delta W}{\delta \rho _{2}} -W,\quad
W=W(\rho _{2},\overset{\cdot }{\rho }_{2}) \label{equationstate}
\end{equation}
plays the Rayleigh-Lamb equation which governs the radial
oscillations of a spherical bubble:
\begin{equation}
R\overset{\cdot \cdot }{R}+\dfrac{3}{2}\overset{\cdot
}{R}^{2}=\dfrac{1}{\rho _{l}}(p_{g}-p_2),\quad
p_{g}=\rho_{g}^2\frac{d\varepsilon_{g}}{d\rho_{g}} \label{RL}
\end{equation}
The equivalence between (\ref{RL}) and (\ref{equationstate}) with
$W$ given by (\ref{potential}) has been proved, for example,  in
Gavrilyuk (1994). Dissipation-free model
(\ref{bubbly})-(\ref{potential}) assumes that the sliding between
components is negligible. Moreover, this model is valid only when
the volume fraction of bubbles is very small. Notice that a
detailed description of the  bubble interaction has been recently
done by Russo \& Smereka (1996) and  Herrero, Lucquin-Desreux \&
Perthame (1999) in the case of rigid bubbles, and  Smereka (2002),
Teshukov \& Gavrilyuk (2002) in the case of compressible bubbles,
by using a kinetic approach.\newline At the rigid walls $y=0$ and
$y=H_{0}$, the vertical components of the velocity are equal to
zero
\begin{equation*}
v_{1}(t,x,0) = v_{2}(t,x,H_{0})=0.
\end{equation*}
In accordance with the previous Section, we prescribe the
following boundary conditions at the contact interface $y=h(t,x)$:
\begin{equation}
h_{t}+u_{1}h_{x}=v_{1},\quad h_{t}+u_{2}h_{x}=v_{2},\quad
p_{1}=p_{2} \label{bc}
\end{equation}
\section{Linear stability problem}
\setcounter{equation}{0}
\subsection{Linearization}
Consider the following main parallel flow of two fluids:
\begin{equation}
u_{1}=u_{10}=const,\quad v_{1}=0,\quad p_{1}=p_{0}=const,\quad
0<y<h_{0},\quad h_{0}=const  \label{mf1}
\end{equation}
\begin{equation}
u_{2}=u_{20}=const,\quad v_{2}=0,\quad p_{2}=p_{0}=const,\quad
\rho _{2}=\rho _{20}=const,\quad h_{0}<y<H_{0}  \label{mf2}
\end{equation}
Small perturbations denoted by $'$\ :
\begin{equation*}
u_{1}=u_{10}+u_{1}^{^{\prime }},\quad v_{1}=v_{1}^{^{\prime }},\quad
p_{1}=p_{0}+p_{1}^{^{\prime }},
\end{equation*}
\begin{equation*}
\rho _{2}=\rho _{20}+\rho _{2}^{^{\prime }},\quad
u_{2}=u_{20}+u_{2}^{^{\prime }},\quad v_{2}=v_{2}^{^{\prime }},\quad
p_{2}=p_{0}+p_{2}^{^{\prime }}
\end{equation*}
satisfy the linearized system of equations
\begin{equation}
u_{1x}^{^{\prime }}+v_{1y}^{^{\prime }}=0,\quad \rho
_{10}D_{1}u_{1}^{^{\prime }}+p_{1x}^{^{\prime }}=0,\;\rho
_{10}D_{1}v_{1}^{^{\prime }}+p_{1y}^{^{\prime }}=0,\quad 0\leq y\leq h_{0}
\label{linpure}
\end{equation}
\begin{equation}
D_{2}\rho _{2}^{^{\prime }}+\rho _{20}(u_{2x}^{^{\prime
}}+v_{2y}^{^{\prime }})=0,\quad \rho _{20}D_{2}u_{2}^{^{\prime
}}+p_{2x}^{^{\prime }}=0,\;\rho _{20}D_{2}v^{^{\prime
}}+p_{2y}^{^{\prime }}=0,\quad h_{0}\leq y\leq H_{0}.
\label{linbulles}
\end{equation}
We use the notations
\begin{equation*}
D_{i}=\frac{\partial }{\partial t}+u_{0i}\frac{\partial }{\partial
x},\quad i=1,2
\end{equation*}
In (\ref{linbulles})
\begin{equation*}
p_{2}^{^{\prime }}=a^{2}\rho _{2}^{^{\prime }}+b^{2}D_{2}^{2}\rho
_{2}^{^{\prime }}
\end{equation*}
with
\begin{equation*}
a^{2}=\rho _{20}\frac{\partial ^{2}W}{\partial \rho _{2}^{2}}(\rho
_{20},0),\quad b^{2}=-\rho _{20}\frac{\partial ^{2}W}{\partial
\overset{\cdot }{\rho }_{2}^{2}}(\rho _{20},0).
\end{equation*}
Here $a$ is the {\it equilibrium} sound velocity, and $b$ is a
characteristic wave length depending  on the bubble radius and gas
volume fraction. We suppose that
\begin{equation*}
\frac{\partial ^{2}W}{\partial \rho _{2}\partial \overset{\cdot
}{\rho }_{2}}(\rho _{20},0)=0
\end{equation*}
This condition is obviously fulfilled for potential
({\ref{potential}}). The coefficients $a$ and $b$ calculated in
equilibrium $R=R_0$ and $p_2=p_0$ are :
\begin{equation}
a^{2}=-\frac{dp_{g}}{d\tau }(\tau_0)\frac{1}{N_0\rho_{20}},\quad
b^{2}=\frac{1}{4\pi N_0 R_0}\frac{\rho_{l}}{\rho_{20}}
\label{coeff}
\end{equation}
For dilute mixtures we can simplify these expressions by replacing
$\rho_{20}$ by $\rho_l$:
$$
a^{2}=-\frac{dp_{g}}{d\tau }(\tau_0)\frac{1}{N_0\rho_{l}},\quad
b^{2}=\frac{1}{4\pi N_0 R_0}=\frac{R_0^2}{3\alpha_0}
$$
Here $N_0=\rho_{20}n$ is the number of bubbles per unit volume,
$R=R_{0}$ is the equilibrium bubble radius, $\alpha_0 =\tau_0 N_0$
is the volume fraction of gas, $\tau_0 =\dfrac{4}{3}\pi R_0^{3}$
is the bubble volume \ and $p_{g}=p_{g}(\tau )$ is the gas
pressure in bubbles expressed as a function of bubble volume. The
equilibrium sound speed $a$ is usually small with respect to the
gas sound velocity. Expressions (\ref{coeff}) can   be obtained
directly after linearization of Rayleigh-Lamb's equation for
bubbles (\ref{RL}).\newline Boundary conditions (\ref{bc}) at
$y=h_{0}$ are
\begin{equation}
h_{t}^{^{\prime }}+u_{10}h_{x}^{^{\prime }}=v_{1}^{^{\prime }},\quad
h_{t}^{^{\prime }}+u_{20}h_{x}^{^{\prime }}=v_{2}^{^{\prime }},\quad
p_{1}^{^{\prime }}=p_{2}^{^{\prime }}  \label{bclin}
\end{equation}
Let us consider the normal modes of the linear problem
(\ref{linpure}), (\ref{linbulles}), (\ref{bclin}):
\begin{equation}
u_{1}^{^{\prime }}=U_{1}(y)\exp (i(kx-\omega t)),\quad v_{1}^{^{\prime
}}=ikV_{1}(y)\exp (i(kx-\omega t)),  \label{ex}
\end{equation}
\begin{equation*}
p_{1}^{^{\prime }}=P_{1}(y)\exp (i(kx-\omega t)),\quad h^{^{\prime }}=H\exp
(i(kx-\omega t))
\end{equation*}
\begin{equation*}
\rho _{2}^{^{\prime }}=R_{2}(y)\exp (i(kx-\omega t)),\quad u_{2}^{^{\prime
}}=U_{2}(y)\exp (i(kx-\omega t)),
\end{equation*}
\begin{equation*}
v_{2}^{^{\prime }}=ikV_{2}(y)\exp (i(kx-\omega t)),\quad p_{2}^{^{\prime
}}=P_{2}(y)\exp (i(kx-\omega t))
\end{equation*}
By substituting into the linearized system we get the system of equations
for unknown amplitudes
\begin{equation}
U_{1}+V_{1y}=0,\quad \rho _{10}(u_{10}-c)U_{1}+P_{1}=0,  \label{amplit}
\end{equation}
\begin{equation*}
-\rho _{10}k^{2}(u_{10}-c)V_{1}+P_{1y}=0,
\end{equation*}
\begin{equation*}
(u_{20}-c)R_{2}+\rho _{20}(U_{2}+V_{2y})=0,\quad \rho
_{20}(u_{20}-c)U_{2}+P_{2}=0,
\end{equation*}
\begin{equation*}
-\rho _{20}k^{2}(u_{20}-c)V_{2}+P_{2y}=0,\quad P_{2}=\left(
a^{2}-b^{2}k^{2}(u_{20}-c)^{2}\right) R_{2}
\end{equation*}
with the following boundary conditions
\begin{equation*}
V_{1}(0)=0,\quad V_{2}(H_{0})=0,\quad P_{1}(H_{0})=P_{2}(H_{0}),\quad \frac{%
V_{1}(h_{0})}{u_{10}-c}=\frac{V_{2}(h_{0})}{u_{20}-c}=H
\end{equation*}
Here $c=\dfrac{\omega }{k}$ is the phase velocity.
\newline From
equations (\ref{amplit}) we obtain the eigenvalue problem for
pressure amplitudes:
\begin{equation}
P_{1yy}-k^{2}P_{1}=0,\quad 0<y<h_{0},  \label{eqp}
\end{equation}
\begin{equation*}
P_{2yy}-\lambda ^{2}k^{2}P_{2}=0,\quad h_{0}<y<H_{0},
\end{equation*}
\begin{equation*}
P_{1}(h_{0})=P_{2}(h_{0}),\quad \frac{P_{1y}(h_{0})}{\rho _{10}(u_{10}-c)^{2}%
}=\frac{P_{2y}(h_{0})}{\rho _{20}(u_{20}-c)^{2}},\quad
P_{1y}(0)=P_{2y}(H_{0})=0
\end{equation*}
where
\begin{equation*}
\lambda ^{2}=1-\frac{(u_{20}-c)^{2}}{a^{2}-b^{2}k^{2}(u_{20}-c)^{2}}.
\end{equation*}
It follows from (\ref{eqp}) that the eigenvalues $c$ are solutions of the
equation
\begin{equation}
\frac{th(kh_{0})}{\rho _{10}(u_{10}-c)^{2}}+\frac{\lambda\
th(\lambda k\left( H_{0}-h_{0}\right) )}{\rho
_{20}(u_{20}-c)^{2}}=0  \label{disp}
\end{equation}
Next we assume that both layers are thin: $k\left(
H_{0}-h_{0}\right) \ll 1$, $kh_{0}\ll 1,$ and dispersion relation
(\ref{disp}) is simplified into
\begin{equation}
\frac{\left( H_{0}-h_{0}\right) }{\rho _{20}}\left( \frac{1}{(u_{20}-c)^{2}}-%
\frac{1}{a^{2}-b^{2}k^{2}(u_{20}-c)^{2}}\right) +\frac{h_{0}}{\rho
_{10}(u_{10}-c)^{2}}=0  \label{char}
\end{equation}

\subsection{Study of dispersion relation}

In case $u_{20}=u_{10}$, equation (\ref{char}) reduces to the
quadratic equation
\begin{equation*}
(1+A)a^{2}-\left(\left( 1+A\right) b^{2}k^{2}+1\right)(u_{20}-c)^{2}=0,\text{ with }A=%
\frac{h_{0}\rho _{20}}{(H_{0}-h_{0})\rho _{10}}.
\end{equation*}
which has only real roots. It means that the flows with equal
velocities are stable.\newline Let us consider the general case
when $u_{20}\neq u_{10}.$ Equation (\ref{char}) can be rewritten
as a polynomial of fourth degree. The stability needs that all
four roots are real. To study this problem, it is convenient to
rewrite the dispersion relation (\ref{char}) in the form
\begin{equation}
F(z)+A\equiv 0  \label{polynome}
\end{equation}
with
\begin{equation}
F(z)=(1-Nz)^{2}\left( 1-\frac{d^{2}}{z^{2}-1}\right), \label{pol1}
\end{equation}
\begin{equation*}
z=\frac{ad}{u_{20}-c},\quad d=\frac{1}{bk},\quad
N=\frac{M}{d},\quad M=\frac{u_{20}-u_{10}}{a}
\end{equation*}
Here $M$ is similar to the Mach number, $d$ is the dimensionless
length of perturbation wave. Obviously, $z$ is real if and only if
$c$ is real. The derivative of (\ref {pol1}) is
\begin{equation}
F^{^{\prime }}(z)=-2Nd^{2}(1-Nz)\varphi (z)  \label{der}
\end{equation}
where
\begin{equation}
\varphi (z)=\frac{1}{d^{2}}+\frac{1-N^{-1}z}{(z^{2}-1)^{2}}.  \label{phi}
\end{equation}
Four cases must be considered
\begin{equation}
1^{\circ }\quad 0<N<1,\text{ \ }d^{2}\leq N^{-2}-1 \label{case1}
\end{equation}
\begin{equation}
2^{\circ }\quad 0<N<1,\text{ }d^{2}>N^{-2}-1
\label{case2}
\end{equation}
\begin{equation}
3^{\circ }\quad N>1,\text{
}d^{2}>\frac{16N^{2}}{27}\left(\left(8N^{2}-6\right)\sqrt{1-\frac{
3N^{-2}}{4}}+8N^{2}-9\right) \label{case3}
\end{equation}
\begin{equation}
4^{\circ }\quad N>1,\text{ }d^{2}\leq
\frac{16N^{2}}{27}\left(\left(8N^{2}-6\right)\sqrt{1-
\frac{3N^{-2}}{4}}+8N^{2}-9\right) \label{case4}
\end{equation}
Each case is illustrated in Figures 1-4. For greater convenience,
the roots $z^i$ of (\ref{polynome}) are shown as the intersection
points of the two graphs $\eta =F(z)$ and $\eta =-A$.

The results are the following:

In case $2$ the dispersion relation defined by
(\ref{polynome})-(\ref{pol1}) has four real roots for $A$ not
large.

In case $3$ the dispersion relation has four real roots for
''intermediate'' values of $A$.

In other cases the dispersion relation has two real roots and two
complex roots.

Precise  formulations are given in the following Propositions.

\begin{proposition}
If inequalities  (\ref{case1}) are satisfied,  equation
(\ref{polynome}) has two real roots and two complex roots for any
positive $A$ (Figure 1).
\end{proposition}

\begin{proposition}
If inequalities (\ref{case2}) are satisfied, the equation $\varphi
(z)=0 $ defined by (\ref{phi}) has one root $z_{01}$ on interval
$I_{2}.$ Equation (\ref{polynome}) has four real roots for $0<$
$A\leq -F(z_{01})$ (Figure 2a) and two real and two complex roots
for $A>-F(z_{01})$ (Figure 2b).
\end{proposition}

\begin{proposition}
If inequalities (\ref{case3}) are satisfied, the equation $\varphi
(z)=0$ defined by (\ref{phi}) has two roots $z_{02}$ and $z_{03}$
on interval $I_{2}$. Equation (\ref{polynome}) has four real roots
 for $-F(z_{02})<$ $A\leq -F(z_{03})$ (Figure 3b), two real and two
complex roots for $A<-F(z_{02})$  (Figure 3a) and $A>-F(z_{03})$
(Figure 3c).
\end{proposition}

\begin{proposition}
If inequalities (\ref{case4}) are satisfied,  equation
(\ref{polynome}) has two real roots and two complex roots for any
positive $A$ (Figure 4).
\end{proposition}

The proofs are given in Appendix.

It is convenient to rewrite conditions (\ref{case1})-(\ref{case4})
in terms of $b,$ $k$ and $M$ :
\begin{equation}
1^{\circ }:\text{ }0<M<1,\text{ }k<\frac{\sqrt{1-M^{2}}}{bM}  \label{case11}
\end{equation}
\begin{equation}
2^{\circ }:\text{ }0<M<1,\text{ }\frac{\sqrt{1-M^{2}}}{bM}<\text{ }k<\frac{1%
}{bM}\text{ \ or }M>1,\text{ }k<\frac{1}{bM}  \label{case12}
\end{equation}
\begin{equation}
3^{\circ }:\text{ }\frac{1}{bM}<k<\frac{1}{bd_{0}(M)}  \label{case13}
\end{equation}
\begin{equation}
4^{\circ }:\text{ }\frac{1}{bd_{0}(M)}<k  \label{case14}
\end{equation}
Here $d_{0}(M)$ is the root of the equation $f(d,M)=0$, where
\begin{equation*}
f(d,M)=d^{2}-\frac{16M^{2}}{27d^{2}}\left(\left(8\left(
\frac{M}{d}\right)^{2}-6\right)
\sqrt{1-\frac{3d^{2}}{4M^{2}}}+8\left( \frac{M}{d}\right)
^{2}-9\right)
\end{equation*}
($0<d_{0}(M)<M$). This equation has a unique solution $d=d_{0}(M)$ in a
domain $0<d<M$ because of the properties $f_{d}(d,M)>0,$ $f(d,M)\rightarrow
-\infty $ as $d\rightarrow 0$ and $f(M,M)=M^{2}>0.$

Inequalities (\ref{case11})-({\ref{case14}}) are illustrated in
Figure 5. For any given flow parameters $M$ and $b$, one can find
a sufficiently large wave number $k$ satisfying inequality
(\ref{case14}). It means that the  flow is always unstable with
respect to perturbations with sufficiently short wave lengths (as
it follows from Proposition 4, the dispersion relation has complex
roots). The flow with subsonic relative velocity is stabilized in
an intermediate interval of wave lengths if the pure liquid layer
is thin enough with respect to the bubbly layer. In the case of
supersonic relative velocity a similar stabilization is observed
for long waves (see Proposition 2 and inequalities
(\ref{case12})). For another intermediate interval of wave lengths
(\ref{case13}) the stabilization of perturbations is attained for
layer depths of the same order (see Proposition 3).

\section{Conclusion and discussion}
We have derived from Hamilton's principle of stationary action
governing equations and  boundary conditions at the contact
interfaces in bubbly fluids. It has been shown that the dynamic
condition on the interface reduces to the continuity of the
average pressure. By using the boundary conditions derived, we
have studied the Kelvin-Helmholtz instability of two superposed
layers of a pure incompressible fluid and a bubbly fluid. We have
shown that in contrast to the case of two incompressible fluids
when the instability develops for any length of perturbations (if
the gravity, capillarity or compressibility  are not taken into
account), the presence of bubbles can stabilize the flow in some
range of perturbation wave lengths.

The stabilizing effect is due to the following reason. The
development of classical Kelvin-Helmholtz instability leads to the
appearance of wave-like bulges at the interface between two
fluids. The presence of bubbles in a fluid  (i.e. new interfaces)
permits one to transform a part of the energy responsible for  the
bulge formation into the energy of radial oscillations of bubbles.

The viscosity effect on the flow stabilization is an important
issue. The method of viscous potential flows developed by Joseph
{\it et al} (1999) in the analysis of Rayleigh-Taylor and applied
by Funada \& Joseph (2001) in the analysis of Kelvin-Helmholtz,
could be used here.

{\bf Acknowledgement}

Support of V.M.T. research by project SMASH of INRIA is gratefully
acknowledged, the author is most grateful to H. Guillard and R.
Saurel. The work was also supported in part by grant of INTAS
01-868. We thank  anonymous referees for useful comments and
references.

\section{References}

Drazin, P.G. \& Reid, W.H. 1981 Hydrodynamic Stability (Cambridge University
Press, Cambridge)

Funada, T. \& Joseph, D.D. 2001 Viscous potential flow analysis of
Kelvin-Helmholtz instability in a channel, {\it J. Fluid Mech.}
{\bf 445}, 263--283.

Gavrilyuk, S. 1994 Large amplitude oscillations and their
thermodynamics for continua ''with memory''. \textit{Eur. J.
Mech., B/Fluids} \textbf{13}, 753--764.

Gavrilyuk, S. \& Shugrin, S. 1996 Media with equations of state
that depend on derivatives. \textit{J. Appl. Mech. Techn. Phys.}
\textbf{37}, 179--189.

Gavrilyuk, S. \& Gouin, H. 1999 A new form of governing equations
of fluids arising from Hamilton's principle, \textit{Int. J. Eng.
Sci.} \textbf{37}, 1495--1520.

Gavrilyuk, S. \& Teshukov, V. 2001 Generalized vorticity for
bubbly liquid and dispersive shallow water equations,
\textit{Continuum Mechanics and Thermodynamics} {\bf 13}, 365-382.

Green, A.E., Laws N. \& Naghdi, P.M. 1974 On the theory of water
waves. \textit{Proc. Roy. Soc. London A} \textbf{338}, 43--55.

Green, A.E. \& Naghdi, P.M. 1976 A derivation of equations for
wave propagation in water of variable depth. {\it J. Fluid Mech.}
\textbf{78}, 237--246.

Gouin, H.  \& Gavrilyuk, S. 1999 Wetting problem for
multi-component fluid mixtures, {\it Physica A} {\bf 268},
291--308.

Herrero, H., Lucquin-Desreux, B. \& Perthame, B. 1999 On the
motion of dispersed balls in a potential flow: a kinetic
description of the added mass effect,  {\it SIAM J. Appl. Math.}
{\bf 60}, 61--83.

Iordanski, S.V. 1960 On the equations of motion of the liquid
containing gas bubbles. \textit{Zhurnal Prikladnoj Mekhaniki i
Tekhnitheskoj Fiziki} N3, 102--111 (in Russian)

Joseph, D.D., Belanger, J. \& Beavers, G.S. 1999 Breakup of a
liquid drop suddenly exposed to a high-speed airstream. {\it Int.
J. Multiphase Flow} {\bf 25}, 1263--1303.

Kogarko, B. S. 1961 On the model of cavitating liquid.
\textit{Dokl. AN SSSR} \textbf{137}, 1331--1333 (in Russian).

Russo, G. \& Smereka, P. 1996 Kinetic theory of bubbly flow I:
Collisionless  case, {\it SIAM J. Appl. Math.} {\bf 56}, 327--357.

Seppecher, P.  1989, The limit conditions for a fluid described by
the second gradient theory: the case of capillarity, {\it
C.R.Acad. Sci. Paris, Serie II} {\bf 309}, 497--502 .

Smereka, P.  2002 A Vlasov equation for pressure wave propagation
in bubbly fluids, {J. Fluid Mech.} {\bf 454}, 287--325.

Teshukov, V.M. \& Gavrilyuk, S.L. 2002 Kinetic model for the
motion of compressible bubbles in a perfect fluid, {\it European
J. Mech. B/Fluids} {\bf 21}, 469--491.

van Wijngaarden, L. 1968 On the equations of motion for mixtures
of liquid and gas bubbles. \textit{J. Fluid Mech.} {\bf 33},
465--474.

\section{Appendix}

\setcounter{equation}{0}

Real roots of equation (\ref{polynome}) belong to the intervals
\begin{equation*}
I_{1}:-(1+d^{2})^{1/2}<z<-1,\text{ \ }I_{2}:1<z<(1+d^{2})^{1/2}
\end{equation*}

Assuming that $u_{20}>u_{10}$ and the wave number $k$ is positive,
we see that the relative Mach number $M$ and the constant $d$ are
positive. Equation (\ref{polynome}) has always at least two real
roots for any positive values of $d,A$ and $N$\ because on the
real axis $F(z)$ tends to $-\infty $ as $|z|$ tends to $1+0$, and
$F(z)=0$ for $|z|=(1+d^{2})^{1/2}$. Additional two real roots can
appear in subsets of real axis where $F^{^{\prime }}(z)$ changes
sign.

Notice, that the interval $I_{1}$ contains exactly one root of
(\ref{polynome}) because, obviously, $F^{^{\prime }}(z)<0$ for
$z<-1$ (see (\ref{der})). We prove below that, depending on the
flow parameters, equation (\ref{polynome}) has one or three real
roots on interval $I_{2}.$

First, let us show that $\varphi ^{^{\prime }}(z)>0$ on $I_{2}$ if
$0<N<1.$ If $\sqrt{3}/2< N < 1$, then the derivative
\begin{equation*}
\varphi ^{^{\prime }}(z)=\frac{3z^{2}-4Nz+1}{N(z^{2}-1)^{3}}
\end{equation*}
vanishes at points $z_{1}<z_{2},$ where
\begin{equation}
z_{2,1}=\frac{2N}{3}\left( 1\pm \sqrt{1-\frac{3N^{-2}}{4}}\right) .
\label{roots}
\end{equation}
The inequality $\varphi ^{^{\prime }}(z)>0$ is fulfilled because
in this case the roots $z_{i}$ satisfy the inequalities
$0<z_{1}<z_{2}<1$, and the interval where $\varphi ^{^{\prime
}}(z)<0$ has no common points with $I_{2}$. For $0<N<\sqrt{3}/2$
the roots $z_{i\text{ }}$ of the equation $\varphi ^{^{\prime
}}(z)=0$ are complex and $\varphi ^{^{\prime }}(z)>0$ on $I_{2}.$

\textbf{Proof of Proposition 1.} It follows  from inequalities
(\ref{case1}) that $1-Nz>0$ for $z\in I_{2}.$ Let us show that
$F^{^{\prime }}(z)>0$ on this interval.
It was proved above that $\varphi ^{^{\prime }}(z)>0$ on \ $I_{2}$ if $%
0<N<1. $ Taking into account that
\begin{equation}
\varphi (\sqrt{d^{2}+1})=\frac{\sqrt{d^{2}+1}}{Nd^{4}}\left( N\sqrt{d^{2}+1}%
-1\right) <0  \label{end}
\end{equation}
we obtain inequalities $\varphi (z)<0,$ \ $F^{^{\prime }}(z)>0$ on interval $%
(1,\sqrt{d^{2}+1}).$ Then the equation (\ref{polynome}) has only
one root on $I_{2}.$ Proposition 1 is proved.

\textbf{Proof of Proposition 2. }In this case $N^{-1}\in I_{2}$, $%
F^{^{\prime }}(z)>0$ for $1<z<N^{-1}$ and $F(z)$ has a local maximum at $%
z=N^{-1},$ where $F^{^{\prime }}(N^{-1})=0,$ $F(N^{-1})=0$ .
Taking into account that $\varphi ^{^{\prime }}(z)>0$ , $\varphi
(\sqrt{d^{2}+1})>0$ and ${\displaystyle{\varphi
(N^{-1})=\frac{(d^{2}+1)N^{2}-1}{d^{2}(N^{2}-1)}<0}}$, we see that
there exists the unique point $z_{01}$ on $I_{2},$ $z_{01}\in
(N^{-1},\sqrt{d^{2}+1}),$ such that $\varphi (z_{01})=0,$
$F^{^{\prime
}}(z_{01})=0.$ At this point $F(z)$ takes a local negative minimum: $%
F(z_{01})<0.$ It is obvious  that equation (\ref{polynome}) has
three roots on $I_{2}$ only if $0<A<-F(z_{01})$ (see Figure 2a).
Proposition 2 is proved.

\textbf{Proof of Proposition 3.} When inequalities (\ref{case3})
are satisfied, the function $\varphi (z)$ is positive in a
neighbourhood of the end points of $I_{2}$ ($\varphi
(z)\rightarrow \infty $ as $z\rightarrow 1+0, $ $\varphi
(\sqrt{d^{2}+1})>0$) and $1-Nz<0$ . The function $F(z)$ can be non
monotone only in the case when $\varphi (z)$ takes negative values
and, consequently, $\varphi ^{^{\prime }}(z)$ vanishes at some point of $%
I_{2}.$ It follows from (\ref{roots}) that the roots of the equation $%
\varphi ^{^{\prime }}(z)=0$ are such that $z_{1}<1,$ $z_{2}>1$ for $N>1.$ It
is easy to verify that a stronger inequality $z_{2}>N$ is valid for $N>1$.
Hence only the root $z_{2}$ of the equation $\varphi ^{^{\prime }}(z)=0$ can
belong to $I_{2}.$ The inequalities
\begin{equation}
a)\ d^{2}>z_{2}^{2}-1,\text{\quad }b)\ \varphi (z_{2})=\frac{1}{d^{2}}+\frac{%
1-N^{-1}z_{2}}{(z_{2}^{2}-1)^{2}}<0  \label{ine}
\end{equation}
provide the inclusion $z_{2}\in I_{2}$ and the existence of the
roots $z_{02},$ $z_{03}$ of the equations $\varphi (z_{02})=$\
$F^{^{\prime }}(z_{02})=0,$\ $\varphi (z_{03})=F^{^{\prime
}}(z_{03})=0$ satisfying inequalities $1<$
$z_{02}<z_{03}<\sqrt{d^{2}+1}.$ The function $F(z)$ has a local
maximum at $z_{02}$ and a local minimum at $z_{02}$. It means that
equation (\ref{polynome}) has three roots on $I_{2}$ if and only if $%
-F(z_{03})<A<-F(z_{02})$ (see Figure 3a). For\ $A>-F(z_{02})$ and $%
0<A<-F(z_{03})$ this equation has only one root (see Figures 3b,
3c). Notice that inequality (\ref{ine}a) is a consequence of
(\ref{ine}b) because for $N>1$
\begin{equation}
\frac{(z_{2}^{2}-1)^{2}}{N^{-1}z_{2}-1}>z_{2}^{2}-1  \label{st}
\end{equation}
Using the identity
\begin{equation*}
\frac{(z_{2}^{2}-1)^{2}}{N^{-1}z_{2}-1}=\frac{16
N^{2}}{27}\left(\left(8N^{2}-6\right)\sqrt{1-
\frac{3N^{-2}}{4}}+8N^{2}-9\right)
\end{equation*}
one can show that inequality (\ref{ine}b) is equivalent to the second
inequality in\ (\ref{case3}). Proposition 3 is proved.

\textbf{Proof of Proposition 4}. In accordance with (\ref{st}) two cases are
possible:
\begin{equation}
a)\
\frac{(z_{2}^{2}-1)^{2}}{N^{-1}z_{2}-1}>d^{2}>z_{2}^{2}-1,\text{\quad
}b)\ \frac{(z_{2}^{2}-1)^{2}}{N^{-1}z_{2}-1}>z_{2}^{2}-1>d^{2}
\label{st1}
\end{equation}
Inequalities (\ref{st1}a) are equivalent to the inclusion
$z_{2}\in I_{2}$ (see (\ref{ine})) and the positiveness of minimal
value of $\varphi (z)$ on $I_{2}$. Using the inequalities $\varphi
(z)>0,$ $1-Nz<0$ and (\ref{der}) we show that $F^{^{\prime
}}(z)>0$. Consequently, equation (\ref{polynome}) has only one
root on $I_{2}$.

If inequality (\ref{st1}b) is fulfilled, then $z_{2}\notin I_{2}$
and $\varphi ^{^{\prime }}(z)$ does not change sign on $I_{2}.$
The function $\varphi (z)$ is positive on $I_{2}$, and $F(z)$ is a
monotone function. We see that equation (\ref{polynome}) also has
only one root on $I_{2}$. Proposition 4 is proved.
\begin{figure}[ht]
\begin{center}
\epsfig{file=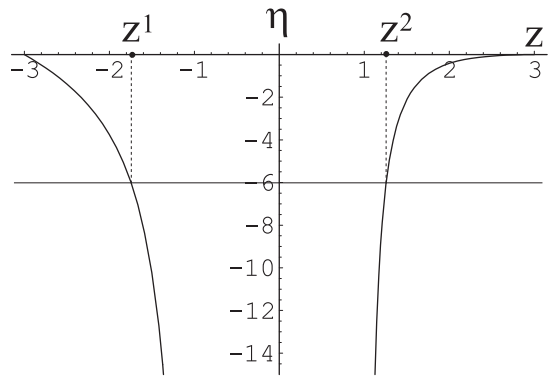}
\end{center}
\caption{The variation of $F$ as a function of $z$ for case 1. The
dispersion relation $F(z) +A =0$ has only two real roots $z^i$,
$i=1,2$ for any positive $A$. Graph is drawn for $N=0.25$ and
$d=2\sqrt{2}$.} \label{Fig1}
\end{figure}

\begin{figure}[ht]
\begin{center}
\epsfig{file=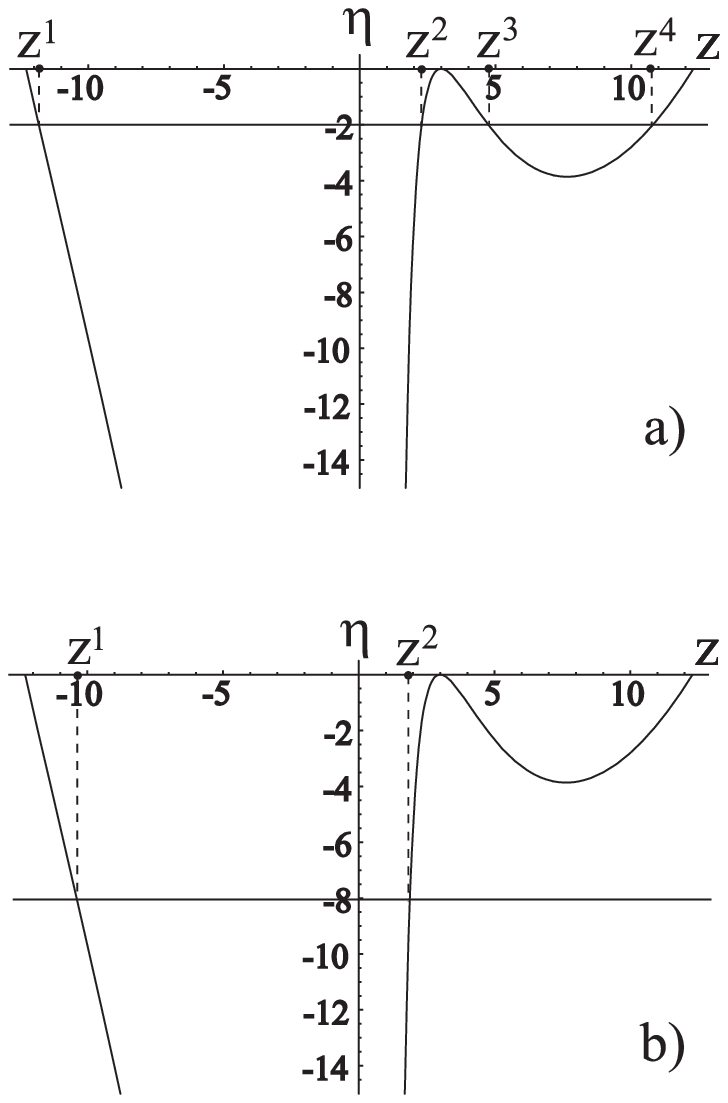}
\end{center}
\caption{The variation of $F$ as a function of $z$ for case 2. The
dispersion relation $F(z) +A =0$ has four real roots $z^i,
i=1,2,3,4$ for not large values of $A$ and two real roots $z^i$,
$i=1,2$ for large values of $A$. Graph is drawn for $N=3$ and
$d=\sqrt{150}$.} \label{Fig2}
\end{figure}

\begin{figure}[ht]
\begin{center}
\epsfig{file=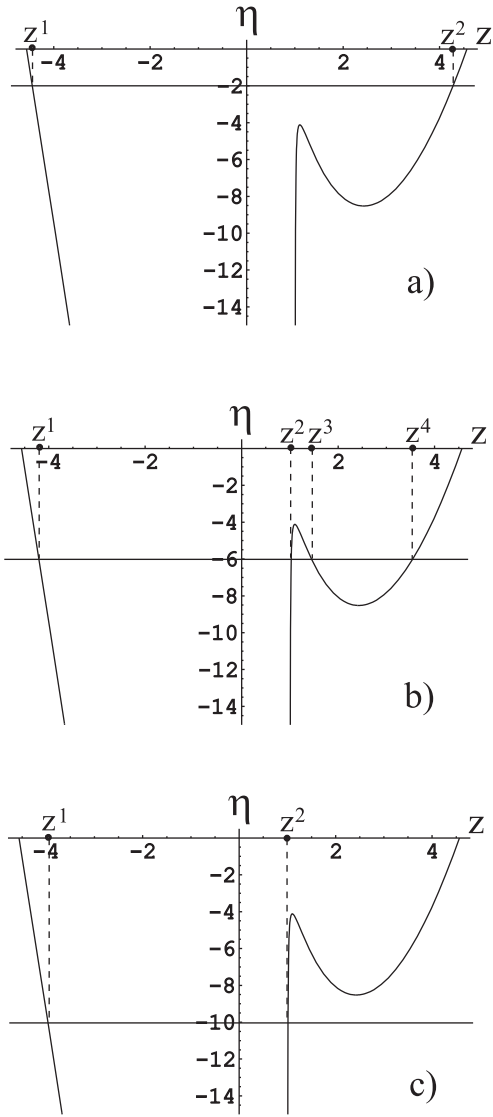}
\end{center}
\caption{The  variation of $F$ as a function of $z$ for case 3.
The dispersion relation $F(z) +A =0$ has  two real roots $z^i$,
$i=1,2$ for $A$ small or large, and four real roots $z^i$,
$i=1,2,3,4$ for $A$ intermediate. Graph is drawn for $N=1.1$ and
$d=4.45$.} \label{Fig3}
\end{figure}

\begin{figure}[ht]
\begin{center}
\epsfig{file=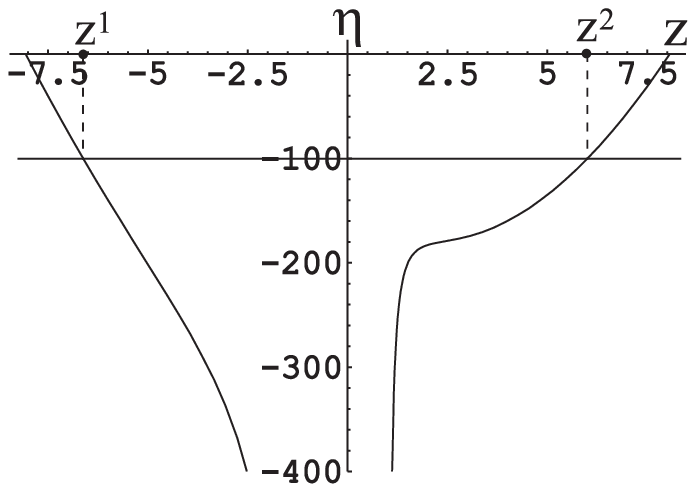}
\end{center}
\caption{The variation of $F$ as a function of $z$ for case 4. The
dispersion relation $F(z) +A =0$ has only two real roots $z^i$,
$i=1,2$ for any positive $A$. Graph is drawn for $N=2$ and
$d=10.5$} \label{Fig4}
\end{figure}

\begin{figure}[ht]
\begin{center}
\epsfig{file=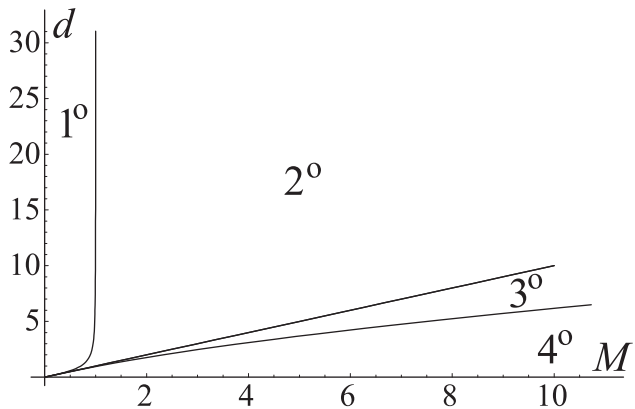}
\end{center}
\caption{Cases 1-4 determined by inequalities
({\ref{case11}})-({\ref{case14}}) define four connected sets in
$(M,d)$-plane. In  regions $2^-$ and $3^-$ four real roots can
exist.} \label{Fig5}
\end{figure}

\end{document}